\documentclass{emulateapj}
\usepackage[utf8]{inputenc}
\usepackage{color}
\usepackage{natbib}
\usepackage[dvipsnames]{xcolor}
\usepackage{graphicx}
\usepackage{amsmath}
\usepackage[T1]{fontenc}
\usepackage{verbatim}

\newcommand{\WF}{{\em WFIRST}}
\newcommand{\kep}{{\em Kepler}}
\newcommand{\KT}{{\em K2}}

\newcommand{\mearth}{{M$_\oplus$}}
\newcommand{\rearth}{{R$_\oplus$}}
\newcommand{\mjup}{{M$_\textrm{Jup}$}}

\newcommand{\msun}{{M$_\odot$}}

\begin{document}

\title{Measuring the Galactic Distribution of Transiting Planets with \WF}
\author{Benjamin T. Montet\altaffilmark{1,4}, 
Jennifer C. Yee\altaffilmark{2,4}, 
Matthew T. Penny\altaffilmark{3,4}}

\email{bmontet@uchicago.edu}

\altaffiltext{1}{Department of Astronomy and Astrophysics, University of Chicago, 
5640 S. Ellis Ave., Chicago, IL 60637, USA}
\altaffiltext{2}{Smithsonian Astrophysical Observatory, 60 Garden
Street, Cambridge, MA 02138, USA}
\altaffiltext{3}{Department of Astronomy, Ohio State University, 140 West
18th Avenue, Columbus, OH 43210, USA}
\altaffiltext{4}{Sagan Fellow}

\date{\today}

\begin{abstract}

    The \WF\ microlensing mission will measure precise light curves
    and relative parallaxes for millions of stars, giving it the potential
    to characterize short-period transiting planets all along the line
    of sight and into the galactic bulge. These light curves
    will enable the detection of more than 100,000 transiting planets
    whose host stars have measured distances. Although most of these
    planets cannot be followed up, several thousand hot Jupiters can
    be confirmed directly by detection of their secondary
    eclipses in the \WF\ data. Additionally, some systems of small
    planets may be confirmed by detecting transit timing variations
    over the duration of the \WF\ microlensing survey. Finally, many
    more planets may be validated by ruling out potential false
    positives. The combination of
    \WF\ transits and microlensing will provide a complete picture of
    planetary system architectures, from the very shortest periods
    to unbound planets, as a function of galactocentric distance.
\end{abstract}
\section{Introduction}
\label{sec:intro}

The \WF\ microlensing survey is designed to detect planets with masses
as small as Mars both bound at separations of several au and free-floating
\citep{Spergel15}. In the course of that 5-year survey, it will give precise photometry of 56 million stars down to $H_{\rm AB}=21.6$ during six 72-day campaigns.
\citet{BennettRhie02}  have suggested that such a microlensing survey has
the potential to detect tens of thousands of transiting giant
planets. \citet{McDonald14} explore the capability of a hypothetical
{\em Euclid} microlensing survey to detect and characterize transiting
planets. In this paper, we expand on the idea of Tanner
\& Bennett in \citet{Spergel15} and consider in detail the ability of
\WF\ to detect transiting planets and to confirm
them directly from the \WF\ microlensing data. Because the majority of bright stars will have
relative parallaxes measured from the \WF\ data \citep{Gould15}, this microlensing survey has powerful implications for the discovery of both transiting and microlensing planets at a wide range of galactic distances \citep{McDonald14}.

To date, all detected planetary systems not found by microlensing have
been found in the local neighborhood, in large part due to
the difficulty of detecting planets farther away.  Most techniques for
detecting planetary systems rely on detecting light from the host
star, which biases them to detections of relatively nearby planetary
systems. For example, radial velocity surveys
\citep[e.g.][]{Udry07, Ford14}, primarily targeted bright, nearby
($\lesssim 100$ pc) FGK stars \citep{Valenti05,Ammons06}.

The \kep\ mission \citep{Borucki10} had the most power to detect
transiting planets farther away, with discoveries out to a kiloparsec
\citep{Lillo-Box14, Barclay15, Quinn15}. Even in this limited volume,
the mission has found some surprising differences between the local
neighborhood and more distant regions of the galaxy.  In particular,
the number of hot Jupiter systems discovered by \kep\ suggests an
occurrence rate approximately 50\% of that suggested by RV detections
of hot Jupiters in the solar neighborhood.  RV surveys estimate an
occurrence rate for hot Jupiters on the order of 1\% \citep{Cumming08,
  Mayor11}, while data from the \kep\ mission suggest an occurrence
rate of $0.4\% \pm 0.1\%$ \citep{Howard12}.  While the difference
between the two fields is known, the explanation is unclear. Studies
have invoked stellar metallicity \citep{Howard12, Wright12, Dawson13},
stellar age \citep{Schlaufman13}, and stellar multiplicity
\citep{Wang14, Wang15a}. Regardless, this result suggests that planet
occurrence rate may be affected by the local galactic environment.

The \KT\ mission is providing the first opportunity to understand the
differences in planet populations across the galaxy. After the failure
of its second reaction wheel, \kep\ became \KT\ and switched to an
observing mode in which it is observing a series of fields in the
ecliptic plane for $\sim 70$ days at a time.  This new mission has led
to catalogs of transiting planet candidates \citep{Foreman-Mackey15,
  Vanderburg16} as well as statistically validated planets
\citep{Montet15b, Crossfield16}.  By the end of the \KT\ mission, it will observe
$\sim 20$ fields, providing an opportunity to probe variations in
planet occurrence along different lines of sight through the galaxy. Nevertheless, because \KT\ is still limited to observations of
bright stars, it will only probe planets within $\sim 1$ kpc of the
Sun.

Previously, OGLE-III \citep{Udalski02b,Udalski02a} conducted searches for transiting planets in microlensing data, 
leading to several detections \citep[e.g. ][]{Dreizler03,Konacki03,Bouchy05}.
However, these detections have been limited to a small number of giant planet
candidates.
The SWEEPS survey also searched for transiting planets towards the bulge,
finding 14 additional giant planet candidates
\citep{Sahu06, Clarkson08}.
Microlensing is often considered the only technique
that offers the opportunity to probe large numbers of planets as far away as the
galactic bulge \citep{Batista14,CalchiNovati15a}. However, as we will
show, the \WF\ microlensing data will enable the detection of potentially tens of thousands of short-period planets
at comparable distances via the transit method. This gives it the opportunity
to make a detailed measurement of the occurrence rate of short-period
planets at a range of galactic distances. Among other things, this could address
the discrepancy between the occurrence rate of hot Jupiters
in the \kep\ field and the solar neighborhood, and if the galactic bulge has a
lower bulk planet occurrence rate than the local neighborhood \citep{Penny16}. Furthermore, as discussed in \citet{McDonald14}, a microlensing survey that detects transiting planets has different selection biases than the \kep\ and \KT\ surveys, which selected a fraction of the stars in the field for which to download postage stamps before observing each field.

The potential of \WF\ to detect large numbers of transiting planets
is complicated by the difficulty of directly confirming those planets
by traditional methods. In general, because the host 
stars of \WF-detected transiting planets will be so faint, it will not
be possible to conduct followup RV observations to confirm their
masses and rule out false positives. However, building on the
experience from \kep\ \citep{Morton14}, we will show that there are
several measurements based on the \WF\ data alone that can be used to
directly confirm or validate these transiting planets.

In this paper, we consider the capability of the upcoming \WF\ mission
to detect, confirm, and characterize transiting planets and the
distribution of those planets in the galaxy. In Section
2, we compare the \WF\ photometry to that of \kep\ and describe
the properties of our injected, detectable planets.  In Section 3, we study the sensitivity of \WF\ to detecting
transit events and project the possible yield for the mission.  In
Section 4, we discuss potential strategies to directly confirm
individual transiting planets discovered by \WF.  In Section 5, we
discuss how systems can be statistically validated by searching for
signatures of false positive events in the data.  In Section 6, we
discuss the galactic distribution of planets uncovered by \WF.  We
conclude in Section 7.

{\section{Simulating \WF\ Transit Detections}
\label{ss:simulations}}

\subsection{Assumed Parameters of the \WF\ Microlensing Survey}

\subsubsection{Survey Duration and Cadence}

Based on the description in \citet{Spergel15}, 
\WF\ will cycle between ten pointings to tile 2.8 square degrees of the sky
towards the galactic bulge.
At each pointing, the telescope will observe for 52 seconds, returning to the
same pointing every 15 minutes.
The microlensing campaign will encompass six 72-day campaigns 
spread over five years.
In this paper, we assume that the observations will be staggered, with three
campaigns each separated by six months at the start of the mission and
three additional campaigns each separated by six months at the end of
the mission.
The exact timing of the campaigns is inconsequential for the search for 
transiting planets, and only slightly affects the search for transit timing 
variations in the data.
We assume that all data will be taken in the W149
band ($\textrm{0.927--2.000}\mu$m) with the exception of one data
point every 12 hours in the Z087 filter ($\textrm{0.760--0.977}\mu$m),
i.e. one Z087 data point for every 47 obtained in W149. The true \WF\ bandpass, W149, is a broadband filter spanning most of the near-IR (0.93-2.00 $\mu m$). In calculating the observed flux for our target
stars we assume W149 = ($J+H+K$)/3; in assuming limb darkening models for each star we take $H$ band as a proxy
for W149.

\subsubsection{Photometric Noise}
\label{sec:kepler}

We consider photometric noise following the standard CCD
signal-to-noise equation. We use the values from the science
definition team (SDT) report \citep{Spergel15} for the photometric
zeropoint of the detector, as well as the bias, read noise, gain, dark
current, and sky brightness.  This report assumes (perhaps conservatively) an error floor in the
photometry of 1 mmag, which we add in quadrature to the calculations
from the SDT. The SDT estimates of the noise are presented in AB magnitudes.

We compare the estimated noise to that of \citet{Gould15}, who
find that, for saturated stars, the precision in a single observation in the
W149 bandpass will scale such that
\begin{equation}
\sigma = 1.0 \times 10^{(2/15)H_{\textrm{Vega}}-2} \textrm{ mmag},
\end{equation}
where $H_\textrm{Vega}$ is the apparent $H$-band magnitude relative to Vega. 
For reference, $H_\textrm{AB} - H_\textrm{Vega} = 1.39$ mag.

Near the saturation limit of 16.1 mag, the SDT estimate of the 
precision is very similar to that of \citet{Gould15}. For significantly
brighter stars, the \citet{Gould15} estimate of the noise is 
markedly lower than the \citep{Spergel15} prescription. 
As saturated stars make up only a small fraction of the
stars in the \WF\ field of view, the choice of noise model does not
appreciably affect our results. 
For consistency, we only consider the \citet{Spergel15} estimate of the noise
throughout this analysis, noting that if the \citet{Gould15} noise estimate
is realized, the performance of \WF\ will be improved at the extreme bright end.

\begin{figure}[htbp!]
\centerline{\includegraphics[width=0.45\textwidth]{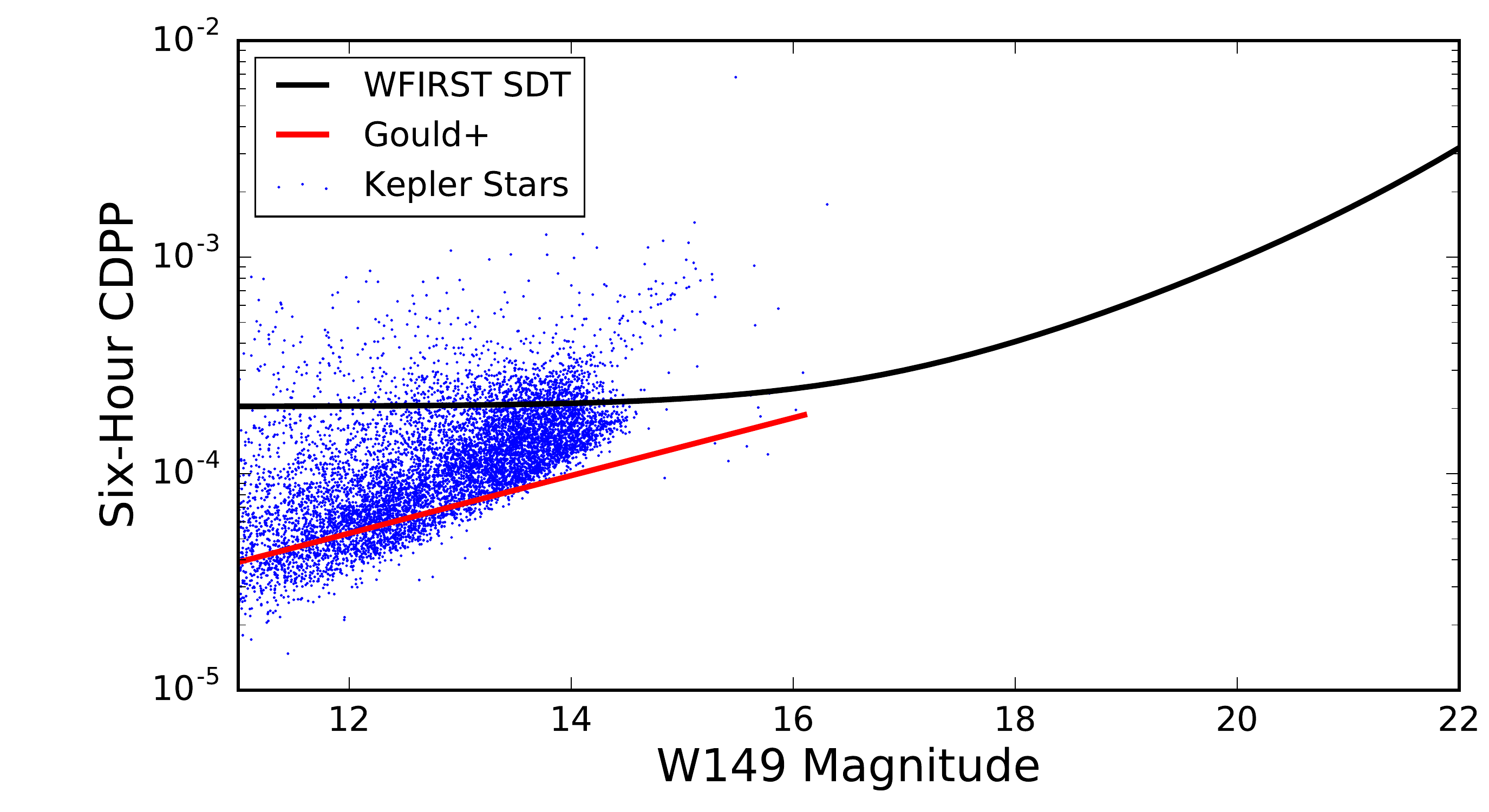}}
\caption{Expected noise properties of \WF\ in the W149
  bandpass as a function of stellar magnitude. The black curve
  represents the estimates of the noise properties from the \WF\ SDT
  report. The red curve represents the estimates of the noise
  properties from \citet{Gould15}, who focus on saturated stars to
  detect asteroseismic modes using \WF\ data.  In blue are actual
  observations of stars from \kep\ for comparison. In all cases, we
  report the six-hour CDPP, or the noise averaged over six hours of
  observations.}
\label{fig:noise}
\end{figure}

We compare both of these relations to the photometric precision of
\kep\ in Figure \ref{fig:noise}.  To perform a direct comparison to
\kep, we make two corrections.  First, we follow the \kep\ convention
of considering the average noise of observations binned over six
hours, the ``combined differential photometric precision'' or CDPP
\citep{Christiansen12}.  Second, the \WF\ bandpass is significantly
redder than the \kep\ bandpass.  As transit searches focus on FGKM
stars, with red colors, these stars appear brighter on the
\WF\ detector than they would on the \kep\ detector.  To provide a
fair comparison, for the \kep\ stars, we use the $H$-band magnitude of
the stars in the \kep\ field.

The expectation is that \WF\ will achieve a relative precision of 1
part per thousand (ppt) in a single observation of a 15th magnitude
star in the W149 bandpass (0.93-2.00 $\mu m$).  This is equivalent to
200 parts per million (ppm) when binned over six hours, comparable to
the precision of \kep\ on a star with $r \approx K_p =15$. However,
since a typical G dwarf has an $R-H$ color of 1.1, the same Sun-like
star observed with \kep\ and \WF\ would be observed at a higher
precision with \WF, even after accounting for the typical extinction
level of $A_H \approx 0.7$ mag toward the bulge.

In this work, we assume the photometric noise is white, so that there
are no correlations between observations.  Correlated noise can be the
result of spacecraft systematics or stellar p-modes
\citep{Gilliland10, Campante11}.  The timescale for p-modes is
inversely proportional with stellar density: for G dwarfs, the
granulation timescale is approximately five minutes; for M dwarfs, 30
seconds.  Like \kep\ data, for most stars observations will be spaced
widely enough to capture a random phase of p-mode oscillations during
each observation.  As \WF\ has significantly larger levels of photon
noise, the correlated stellar signals will be small by comparison,
causing the while instrumental noise to dominate over any red
astrophysical effects.
Moreover, as \WF\ observes at much redder wavelengths, the complicating effects 
of stellar spots will be diminished.

\subsection{Simulated Light Curves}

We simulate individual transits by injecting a planetary signal into
simulated \WF\ data using the prescription for the photometric noise
described in the previous section.  Every fifteen minutes, starting at
a random phase, we collect an observation of the flux from this
system: every twelve hours one observation is taken in Z087, while all
other observations are in W149.  We model the transit light curve with
the transit model of \citet{Mandel02}.  We calculate limb darkening
coefficients in each bandpass using the online tool developed and
described in \citet{Eastman13}, which interpolates the
\citet{Claret11} quadratic limb darkening tables.

Note that unlike the \kep\ mission, each data point will consist of a
single 52-second observation (290 s at Z087), rather than a series of binned
observations over 30 minutes.  Each observation will then sample one
specific point on the transit light curve as opposed to an integrated
measure of the observed flux, meaning morphological light curve
distortions due to finite integration time will be minor in the \WF\ data \citep{Kipping10}.

\subsection{Simulated Host Star Population}

To simulate a realistic estimate of the stellar population in the
\WF\ microlensing field, we develop a galactic population generated from the online
Besan\c{c}on models of the galaxy \citep{Robin03}.  To convert the
returned apparent magnitudes to near-infrared simulated photometry, we
apply the transformations of \citet{Bilir08}.  We then apply a
correction for interstellar extinction assuming the \citet{Cardelli89}
extinction law with $R_v = 2.5$, following \citet{Nataf13}.  From the
derived JHK magnitudes, we approximate the W149 magnitude for each
star by assuming W149 = (J+H+K)/3.

We then apply a series of corrections to turn the Beasan\c{c}on models
into a realistic simulation of the stars observed by \WF. The
Beasan\c{c}on model outputs the properties and numbers of stars along
a given sightline within a certain solid angle.  Because each
simulated field is not a perfect match to the \WF\ field, we weight
each simulated star by the fraction of the simulated field that falls
in the \WF\ field.  We then apply a correction for the mass function
in the bulge.  The model assumes stars in the bulge follow the
Salpeter IMF \citep{Salpeter55}.  We downweight stars of mass $M< 0.5$
\msun\ by a factor of $0.5/M$, which approximates the IMF of
\citet{Kroupa01}.  We then apply a uniform correction to all stars to
match the overall number of bulge main sequence stars near the
\WF\ fields as measured by \citet{Calamida15}. Further details can be
found in \citet{Penny16b}.

\subsection{Simulated Planet Population}

We simulate two planet populations. First, we simulate planets
assuming the occurrence rate is the same as for the \kep\ field. We
assign planets around solar-type FGK stars following the planet
occurrence estimates of \citet{Howard12}.  We assign planet radii and
orbital periods following the ``Cutoff Power-Law Model'' of Table 5 of
that paper, and bulk occurrence rates for each spectral type following
the authors' Table 4.  For M dwarfs, we follow the relations of
\citet{Morton14}, specifically their ``logflat+exponential'' model of
the period distribution from their Figure 7 and the radius
distribution from their Figure 6.  This leads to considerably smaller
numbers of giant planets injected around M dwarfs than more massive
stars, in line with observations from \kep.

We also simulate planets taking host star metallicity into account.
\WF\ will observe stars at large distances along the galactic
metallicity gradient \citep{Rolleston00, Pedicelli09}.  
Observations suggest the average stellar metallicity changes by -0.05 dex/kpc along a line of sight moving radially outward from the center of the galaxy.
Thus \WF\ is
expected to observe stars at preferentially higher metallicities than
the solar neighborhood.  Indeed, simulations of the \WF\ field suggest
the median G2V dwarf observable by \WF\ with W149 < 19.5 has [Fe/H] =
0.26 (Section \ref{ss:yield}). This is significant because radial velocity
surveys have unveiled a correlation between giant planet occurrence
and stellar metallicity \citep{Fischer05, Johnson10a}. Note, however,
that the presence of small transiting planets does not appear to be
affected by the host star's metallicity \citep{Buchhave15}.

To account for this metallicity effect, we weight the planets based
on their radii and host star metallicity. Following
\citet{Johnson10a}, who find planet occurrence scales as
$10^{1.2\textrm{[Fe/H]}}$, we modify the likelihood of all planets
with radii larger than 5 \rearth\ by this factor.  For the median star
([Fe/H] = 0.25), this factor increases giant planet occurrence by a
factor of two.

\section{Transiting Planet Detection with \WF}

\subsection{Detection of Transit Events}
\label{sec:transits}

After all transits have been simulated, we phase-fold on the known
period and measure the significance of the observed transit depth. We
use the same $7.1\sigma$ threshold as that used by \kep\ to decide
whether or not a transit is detected. We also require at least two
transits during at least one season to be detected. 

The $7.1\sigma$ threshold is not strictly appropriate for these
simulations. It was chosen for \kep\ so that there would not be more
than one false positive with three ``transits'' out of 100,000
stars. \WF\ will observe a thousand times more stars, which would
increase the threshold for detection. At the same time, we have
required 2 transits to be detected in a single season. Since \WF\ will
have six microlensing seasons, this means there will be $\sim 12$
transits over the course of the mission. Therefore, the effective
detection threshold over the mission is higher than
$7.1\sigma$. Finally, $7.1\sigma$ was chosen in the absence of
correlated noise and systematics, so in practice the true threshold
for planet detection in \kep\ is higher because of the presence of
these effects. Thus, we conclude that $7.1\sigma$ is a reasonable
benchmark for the detection of planets, but the true threshold will
have to be evaluated once the properties of the data are better
understood.

\begin{figure*}[htbp!]
\centerline{\includegraphics[trim={0 0.0in 0 0.0in}, clip,width=0.9\textwidth]{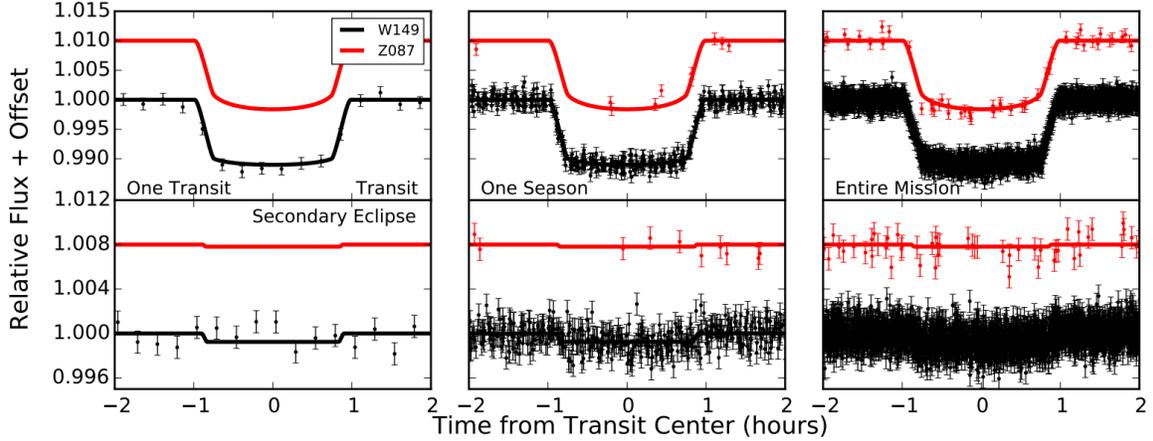}}
\caption{(top) Simulated transit photometry for a hot Jupiter on a
  three-day orbit around a Sun-like star with W149 $= 15$. In black is
  photometry from the W149 bandpass; in red, the Z087 bandpass. The
  left panel corresponds to a single transit. The middle panel
  corresponds to transits folded together over a 72-day observing
  season, while the right panel corresponds to six such seasons over
  the course of the mission. The transit simulated in all panels are identical.
  (bottom) Simulated secondary eclipses for
  the same planet, assuming its planet has an equilibrium temperature
  of 900 K.  The secondary eclipse is then as deep as the transit of a
  3 \rearth\ planet across the same star and is detected at high
  significance by the end of the mission.}
\label{fig:HJtrans}
\end{figure*}

Figure \ref{fig:HJtrans} shows an example light curve for a
Jupiter-sized planet on a 3.0 day orbit around a W149 $=15.0$ mag host
star with impact parameter $b=0.5$. The transit duration is
approximately two hours. These transits can be seen by eye, even in
the case of single transit events.  Over the course of the mission,
more than 150 transits of such a hot Jupiter would be observed,
leading to approximately 1200 observations during the transit in the
W149 bandpass.  Moreover, approximately two dozen observations during
the transit will be collected in the Z087 bandpass, which might be
useful for confirmation of the planetary nature of this signal
(Section \ref{sec:confirm}). By fitting
transit models and evaluating their likelihood, we measure a transit
depth of $0.998 \pm 0.002$ $R_J$, assuming perfect knowledge of the
stellar host. This planet is detected at $\sim 500 \sigma$.

\begin{figure*}[htbp!]
\centerline{\includegraphics[width=0.9\textwidth]{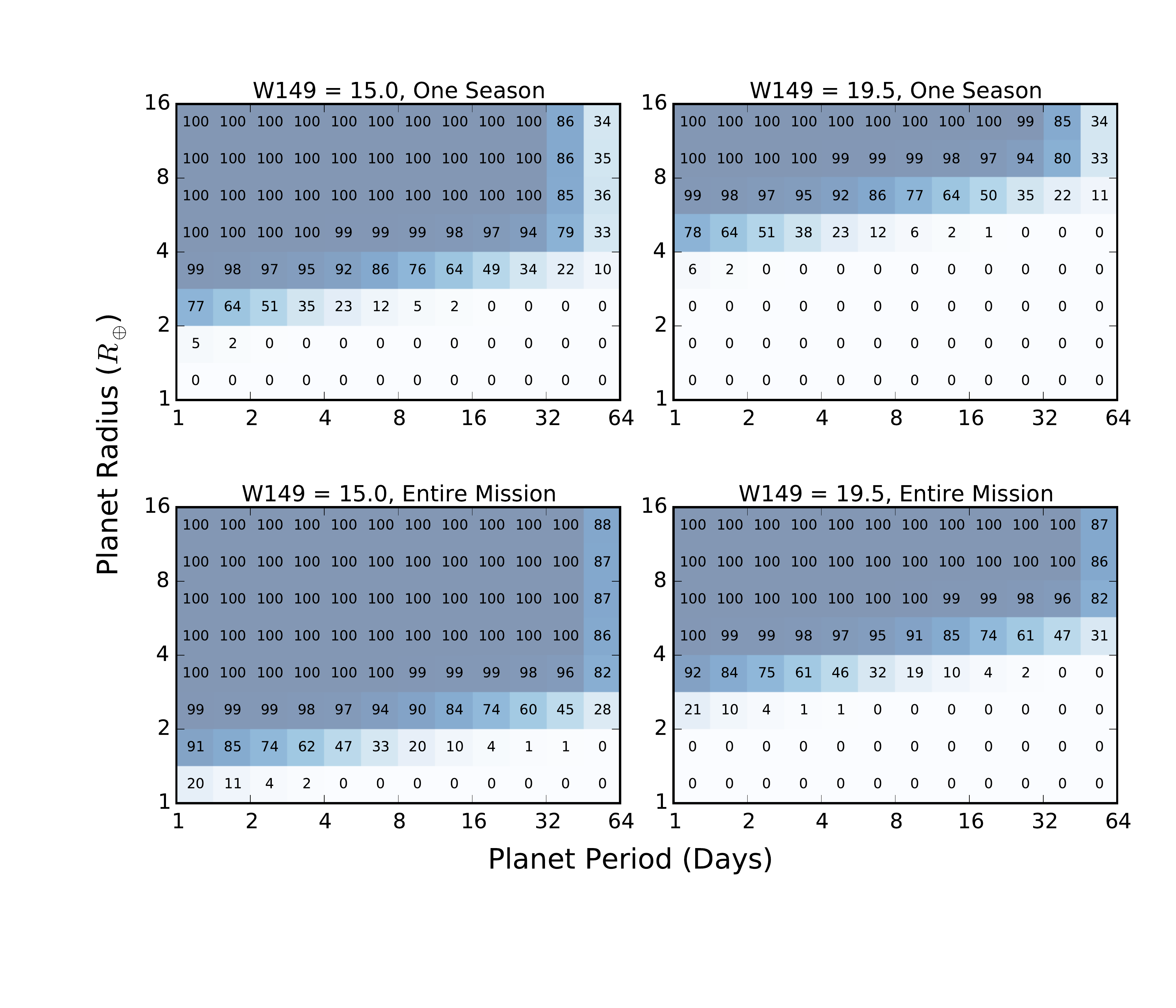}}
\caption{Detectability of planets transiting a Sun-like star in
  simulated \WF\ data by analyzing (top) one season of data and
  (bottom) data from the entire mission.  Around very bright stars
  (W149 = 15.0) nearly all Neptune-sized planets and larger with
  orbital periods shorter than the seasonal baseline will be detected
  in a single season of data.  To qualify as a detection, we require
  at least two transits in a single observing season, but not
  necessarily in all seasons.  }
\label{fig:Sensitivity}
\end{figure*}

Using our simulated light curves we calculate \WF's sensitivity to planets
as a function of radius and period. We simulate
planets with radius and orbital period drawn from log-flat
distributions over the ranges $[1, 16]$ $R_{\oplus}$ and $[1,72]$ days,
respectively. We then assign an impact parameter for each simulated
transiting planet drawn from a uniform distribution over the range
$[0, 1]$. We assume the host star is a G-dwarf with radius $1 R_{\odot}$. We then check to see which planets meet our detection threshold. The results are shown in Figure \ref{fig:Sensitivity}.  We have chosen to present this calculation in terms of physical parameters rather than transit depth because they are more intuitive. However, since the radius of the host star is fixed, it is trivial to convert to transit depth if desired.

We find that, for the brightest stars observed by \WF, Neptune-sized
planets with orbital periods shorter than one month will be easily
detected in a single season of data.  The mission will also recover
many mini-Neptunes with periods shorter than 20 days, and is likely to
recover a small number of planets smaller than 2 \rearth\ with periods
shorter than two days.  Over the entire mission, \WF\ will be
sensitive to a few Earth-sized planets with orbital periods shorter
than two days orbiting the brightest stars.  Of the 12 million stars
with $\textrm{W149} < 19.5$, the prospects for detecting super-Earths
or mini-Neptunes are much lower, but the mission will detect the
majority of Neptune-sized planets with periods less than a month and
all transiting Jupiter-sized planets in that period range as well. We
discuss expected planet yields in Section \ref{ss:yield}.

\subsection{Expected Planet Yield}
\label{ss:yield}

We use the simulations described in Section \ref{ss:simulations} to calculate the number of transiting planets that will be detected by \WF.
We inject planets around the main-sequence dwarf stars brighter
than W149 = 21.0. We use the same detection criteria as in Section \ref{sec:transits}, and thus, we limit the range of orbital periods to $P<72\,$ days.

The results are shown in the left-hand panels of Figure
4. Assuming a \kep-like planet population, we expect \WF\ to detect approximately 13,000
transiting planets orbiting dwarf stars with W149 < 19.5, the majority
being giant planets orbiting F and G stars.  Similarly, we expect
\WF\ to detect 70,000 transiting planets orbiting stars with W149 <
21.0.  The mission will also detect approximately 800 planets smaller than
Neptune, the majority of which will be orbiting M dwarfs. While large, these numbers are not unexpected given that \WF\ will observe two orders of magnitude more stars than \kep, which detected several thousand transiting planets.

\begin{figure*}[!tbp]
  \centering
  \begin{minipage}[b]{0.48\textwidth}
    \includegraphics[width=\textwidth]{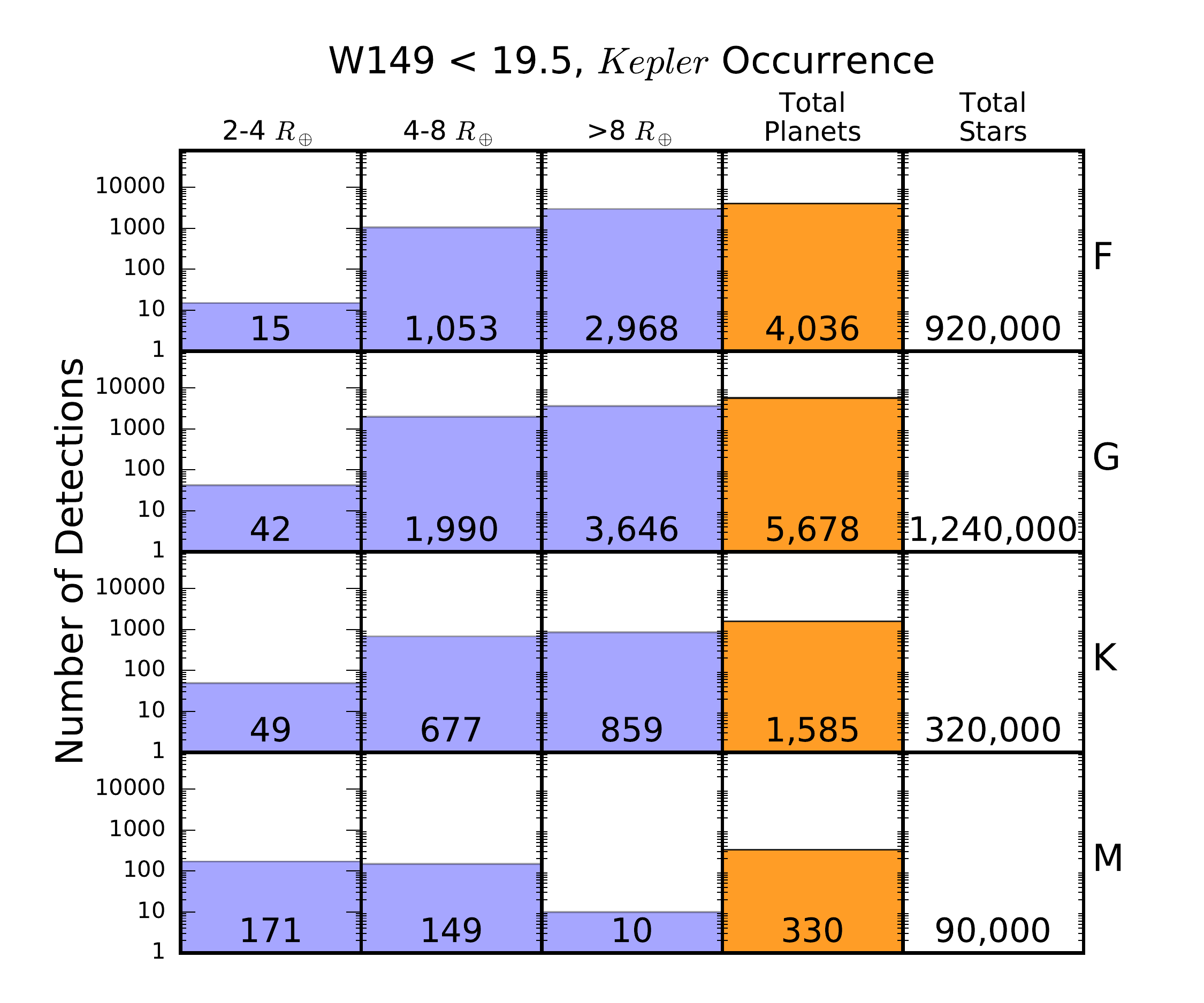}
  \end{minipage}
  \hfill
  \begin{minipage}[b]{0.48\textwidth}
    \includegraphics[width=\textwidth]{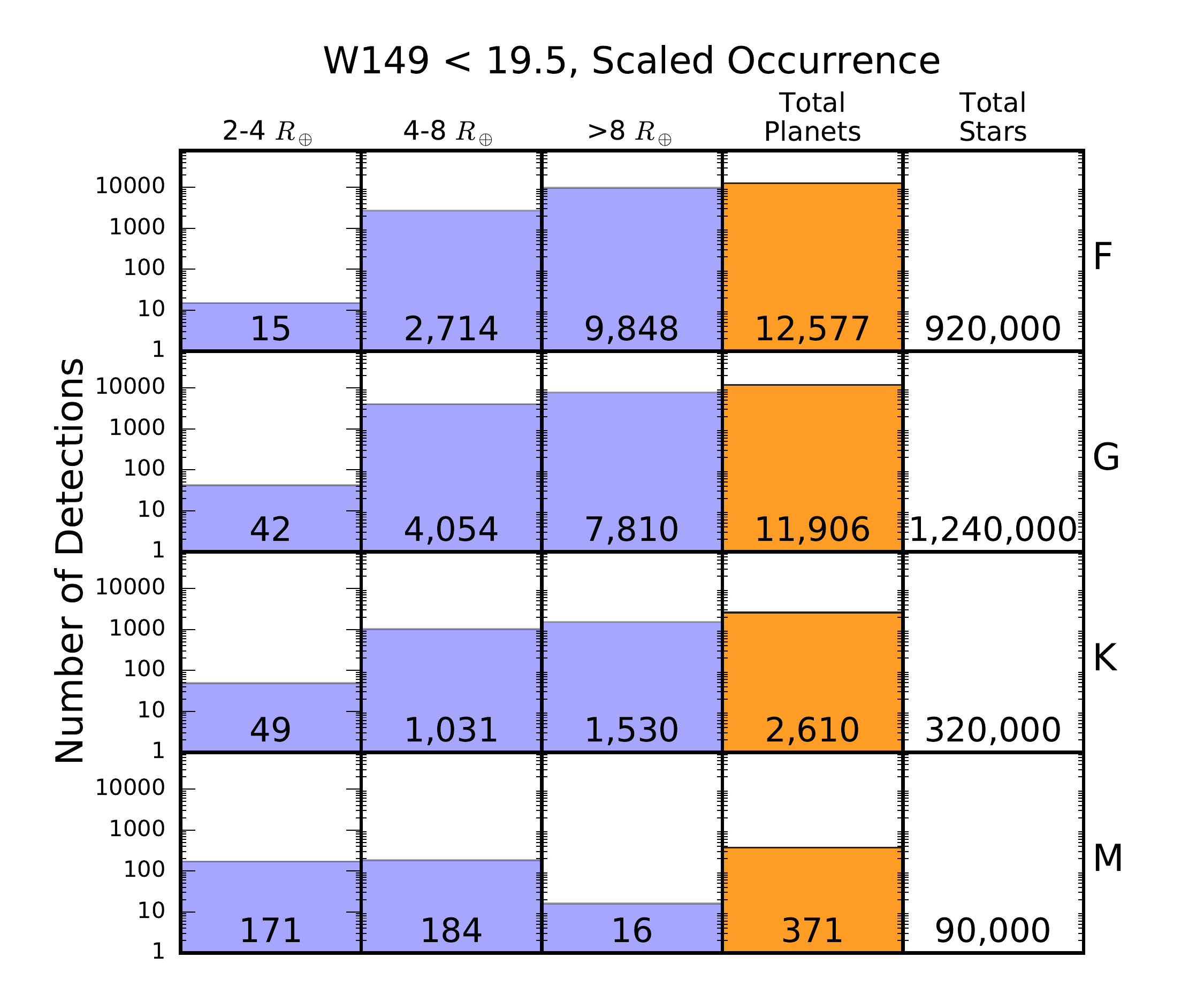}
  \end{minipage}
  \\
    \begin{minipage}[b]{0.48\textwidth}
    \includegraphics[width=\textwidth]{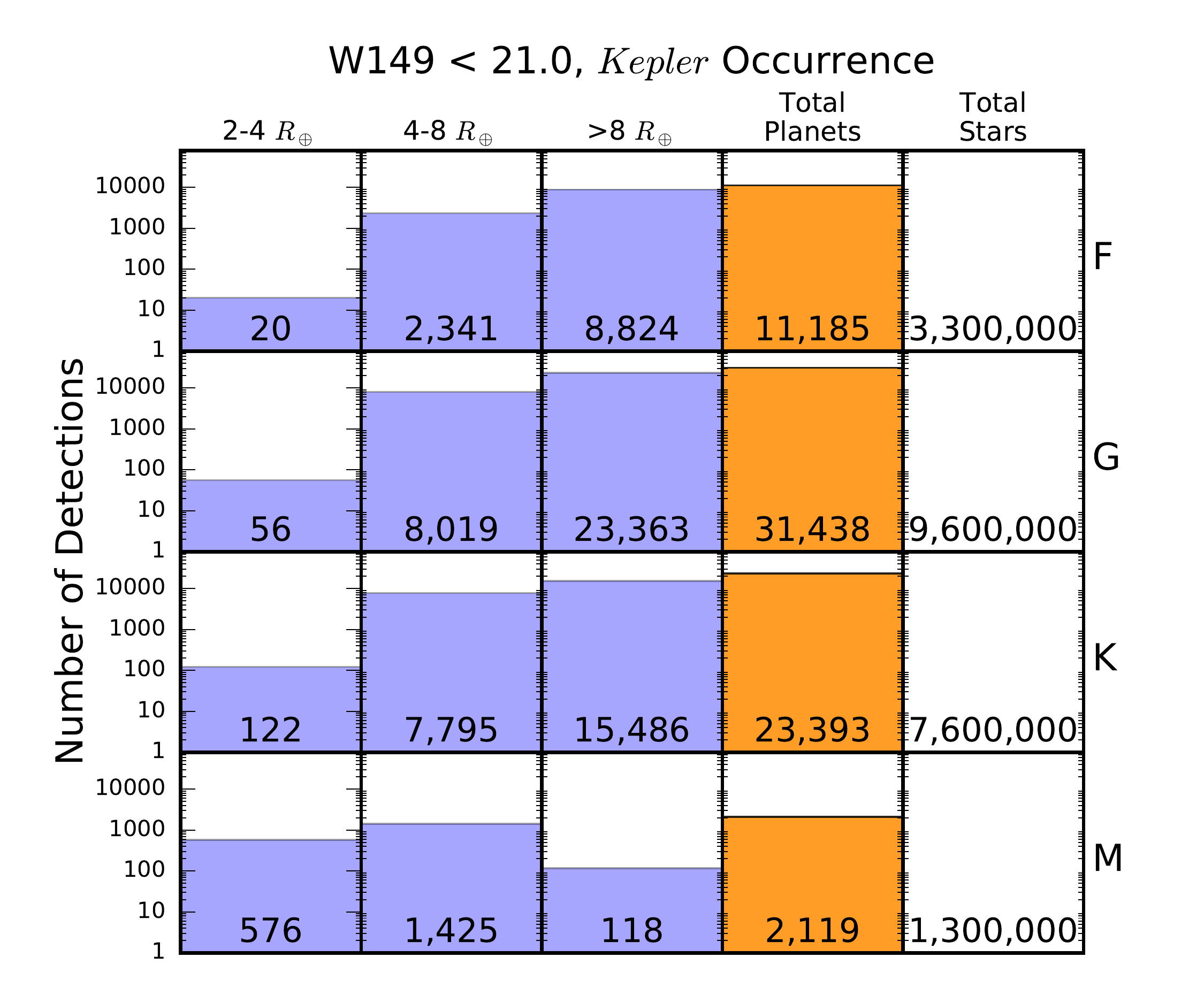}
  \end{minipage}
  \hfill
  \begin{minipage}[b]{0.48\textwidth}
    \includegraphics[width=\textwidth]{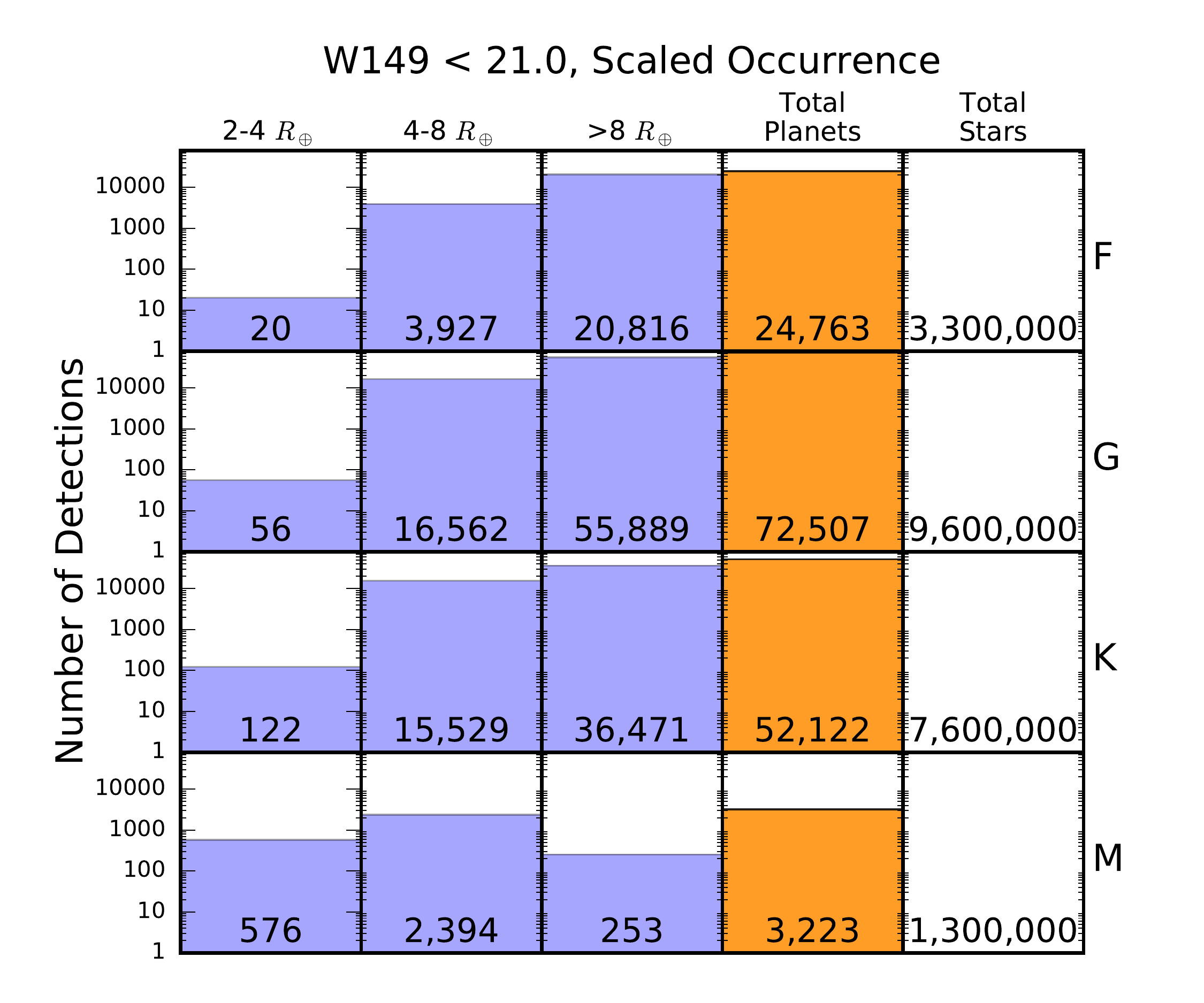}
  \end{minipage}
  \label{fig:yield}
  \caption{(Top) (Left) Expected yield of transiting planets orbiting
    dwarf stars brighter than W149 = 19.5 in the \WF\ data as a
    function of planet size and stellar type, assuming the planet
    occurrence is the same as that in the \kep\ field. \WF\ will
    detect thousands of Jupiter sized planets, but also more than 100
    planets smaller than Neptune, mainly around M dwarfs. (Right) The
    same, but assuming the occurrence rate of planets larger than 5
    \rearth\ follows the metallicity relation of \citet{Johnson10a}.
    (Bottom) Same as the top, with a limiting magnitude of W149=21.0.
    In this case, more than 150,000 transiting planets could be
    detected by the end of the mission.}
\end{figure*}

\begin{figure}[htbp!]
\centerline{\includegraphics[width=0.45\textwidth]{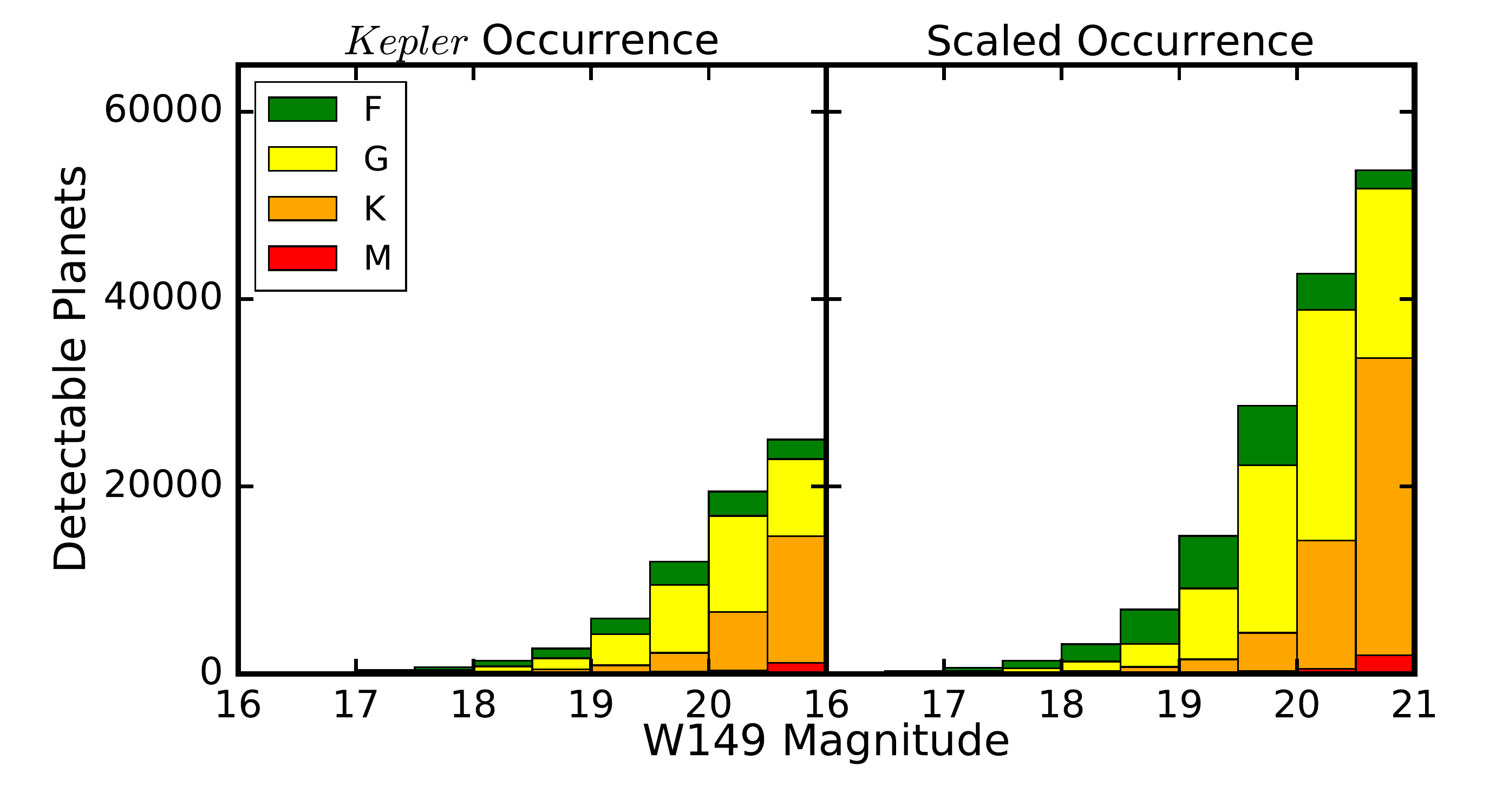}}
\caption{Distribution of apparent magnitudes of the host stars of
  planets detectable by \WF. Similar to the \kep\ mission, the vast
  majority of planets will be found around relatively faint stars.  }
\label{fig:mags}
\end{figure}

The numbers are even more striking when taking into account the expected metallicity dependence on the occurrence rates of transiting planets (right-hand panels of Figure \ref{fig:yield}).  In this case, we detect more
than 150,000 transiting planets over the six seasons of the
\WF\ mission.  As expected, the number of small planets is unchanged,
with the gains made entirely in the population of planets larger than
Neptune.  \WF, by completing this survey, will provide the best
assessment of the effects of high metallicity on the population of
giant planets, providing clues to the formation and evolution of these
systems. With the development of multi-object NIR spectrographs for
large telescopes like the VLT, reconnaissance spectroscopy for large numbers
of faint stars to measure metallicities will be possible \citep{Cirasuolo12}. 
We show the distribution of these planets with respect
to the apparent magnitudes of their host stars in the \WF\ bandpass in
Figure \ref{fig:mags}.

\subsection{Other Transiting Planet Detections}

\subsubsection{Single Transit Events}

\WF\ will also be able to detect large numbers of single
transit events. Planets with no more than one transit per season do not meet our detection criteria. Such planets may or may not have a transit in a subsequent season. If a second transit is observed, there will be some ambiguity in the period because the data are not continuous. However, it should be possible to rule out a substantial fraction of the aliases by considering the non-detections and by estimating the period from the transit ingress and egress following \citet{Yee08}. Even for the cases for which only one transit is observed, it should still be possible to place constraints on the period using these methods, and where possible, prompt RV observations would be extremely beneficial \citep{Yee08}.
A few dozen
singly transiting were planets detected in the \kep\ and \KT\ data,
largely through visual inspection \citep{Wang15b,
  Osborn16,Uehara16,Foreman-Mackey16}. The probability of detecting a
singly transiting planet with \WF\ scales as $P^{-5/3}$, so most
detections of single transits will be of planets with shorter orbital
periods. However, given the large number of stars observed by \WF,
there should still be large numbers of planets with periods of a few years.
While the period distribution of expected microlensing planets with \WF\ peaks
at periods of about ten years, there should be a large number of massive
planets with periods 2-5 years detected as well.  
These longer-period planets offer the opportunity for direct comparison
to the \WF\ microlensing planet population, which will have periods of a few years.
They may also be compared to measurements of the
occurrence rates of long-period planets from the combination of
long-term RV accelerations with direct imaging surveys \citep[Gonzales
  et al. in prep]{Montet14}.
  
\subsubsection{Planets Around Evolved Stars}

Finally, we note that our analysis is limited to dwarf stars towards
the bulge brighter than W149 = 21.0. While planets have been detected
around evolved stars in transit \citep{Lillo-Box14, Barclay15, Quinn15},
and through radial velocities \citep{Johnson11b, Otor16}, the
occurrence rate of short-period planets around evolved stars are too poorly
understood to enable a reliable
estimate of their yield in \WF.  However, given the photometric
precision from Section \ref{sec:kepler} and scaling from Section
\ref{sec:transits}, giant planets will be detectable around evolved
stars (i.e. given that $3R_{\oplus}$ planets are detectable around a
$1 R_{\odot}$ star, a $12R_{\oplus}$ planet should be detectable
around a $4 R_{\odot}$ star.).  As \WF\ will observe large numbers of
evolved stars towards the bulge, it will provide the best measurement
to date of the occurrence rate of giant planets in short orbits around
evolved stars.

\section{Confirmation and Characterization of Transiting Planetary Systems}
\label{sec:confirm}

The major challenge for transiting planet studies is to verify that
the observed transiting object is a planet rather than a false
positive.  Multiple astrophysical events can be mistakenly identified
as transiting planets.  First, because of degeneracy pressure,
Jupiters, brown dwarfs, and low-mass M stars all have similar radii
\citep{Chabrier97}.  Therefore, detection of a Jupiter-radius transit
depth alone is insufficient to claim a planetary detection.  Second, a
false positive can occur in the case of blended light, when the star
in question in the aperture is actually the combined light of multiple
stars.  For example, an unresolved, background eclipsing binary could
be blended with the primary target star.  Similarly, the primary
itself could be an eclipsing binary blended with the chance alignment
of a background star or the light of a hierarchical triple third star.
While these degeneracies are easily resolved with RV observations, those
will not be possible for most \WF\ transit candidates.  However,
previous studies have shown in the case of \kep\ that it is possible
to validate transiting planet candidates by ruling out various false
positive scenarios \citep{Morton12, Morton16}. Here we explore various means to confirm  \WF\ transiting planet candidates. In Section
\ref{sec:validate} we consider ways to validate or
rule out false positives for planets that cannot be confirmed directly. See also \citet{McDonald14} for a discussion of these topics with respect to a {\em Euclid} microlensing survey.

\subsection{Multiple Planet Systems}

If \WF\ observes transits from multiple planets around a single star,
this by itself significantly increases the probability that the
transits are indeed due to real planets rather than astrophysical
false positives.  The initial \kep\ data release contained 444
multiple-candidate systems from a total of $\sim 1600$ candidate
systems \citep[see ][]{Lissauer11b,Lissauer12}. Thirty-eight of those
systems have $r>3R_{\oplus}$ and $P<72$ days, making them easily
detectable by \WF. It would not be surprising for \WF\ to discover more than one
thousand planetary systems with multiple transiting planets over the
course of its mission.

\subsection{Transit Timing Variations}
\label{sec:ttvs}

If there are multiple planets in a system, this opens the possibility
of measuring transit timing variations (TTVs).  TTVs are the deviation of the time of the transit center from a linear ephemeris and are caused by dynamical interactions between the various bodies in the system \citep{Holman05,Agol05,LithwickWu12}. The exact nature of any TTV curve depends on the architecture of any
particular TTV system: two planetary systems with identical planets
but different orbital eccentricities, arguments of periapse, or
longitudes of ascending node would exhibit different TTV signals. They are the most straightforward way for \WF\ to
directly confirm the planetary nature of transit signals.

\kep\ has had enormous success at measuring TTV signals. They have been used 
to confirm the planetary nature of transiting signals \citep{Holman10,
  Fabrycky12, Ford12TTV, Xie13}, to detect the presence of
non-transiting planets \citep{Ballard11, Nesvorny12, Nesvorny13}, and
to infer masses and eccentricities of planetary systems
\citep{Hadden14, Jontof-Hutter15, Jontof-Hutter16}.  Based on data
from the \kep\ mission, TTV signals have been analyzed around more
than 2,500 KOIs \citep{Holczer15, Holczer16}.  They have detected 260
KOIs with TTVs on timescales $>100$ days, i.e. likely to be due to a
companion rather than an astrophysical false positive. Of these, 163
have TTV amplitudes larger than 15 minutes (see below). Based on the
number of planets we expect \WF\ to be able to detect and the timing
precision we expect the mission to achieve on individual transits
\ref{sec:ttvs}, hundreds of systems with observable TTVs should be
detected over the observing campaign.

We should note that \WF\ TTVs will have a few differences with respect to \kep\ TTVs. The longer time baseline (5 years as opposed to 4) will
enable the possible detection of transit timing signals with longer
periods, such as those due to the Roemer delay from a hierarchical
binary star or the orbital evolution of a giant planet in a short
orbit \citep{Ragozzine09, Maciejewski16}. However, the
large gaps between seasons result in degeneracies in the TTV solutions. At the same time, \WF\ TTVs will be less
severely affected by starspots. Starspots complicate the measurement of TTVs by distorting the light curve both in and out of transit. Since
\WF\ will observe in the near-IR, where the effects of starspots are
significantly minimized due to their lower contrast, this reduces the
possibility of significant starspot-induced timing errors.

To better understand the detection of TTVs with \WF, we simulate
transit events in order to estimate the precision to which we will be
able to measure transit times. We model our benchmark system after
Kepler-9b and Kepler-9c, the first planets confirmed via TTVs
\citep{Holman10}.  We use orbital periods for the two planets of 19.2
and 38.9 days. \kep\ has shown that less massive planets more often exhibit TTVs
than more massive planets \citep{Mazeh13}, and yet a transit must be
detected in order to measure a TTV. Thus, we simulate planets near the
bottom of our detectability contours in order to understand a typical
TTV signal seen by \WF. We assume the two planets are
mini-Neptunes with masses of $10$ \mearth\ and radii of $3$
\rearth. These are significantly smaller than the real Kepler-9
planets leading to smaller TTVs and larger uncertainty in the measured
time of transit center.  We simulate transits of these planets
orbiting a Sun-like star with W149 = 15.0, so that the photometric
precision on each data point is 1 part per thousand, assigning impact
parameters at random.

\begin{figure}[htbp!]
\centerline{\includegraphics[width=0.45\textwidth]{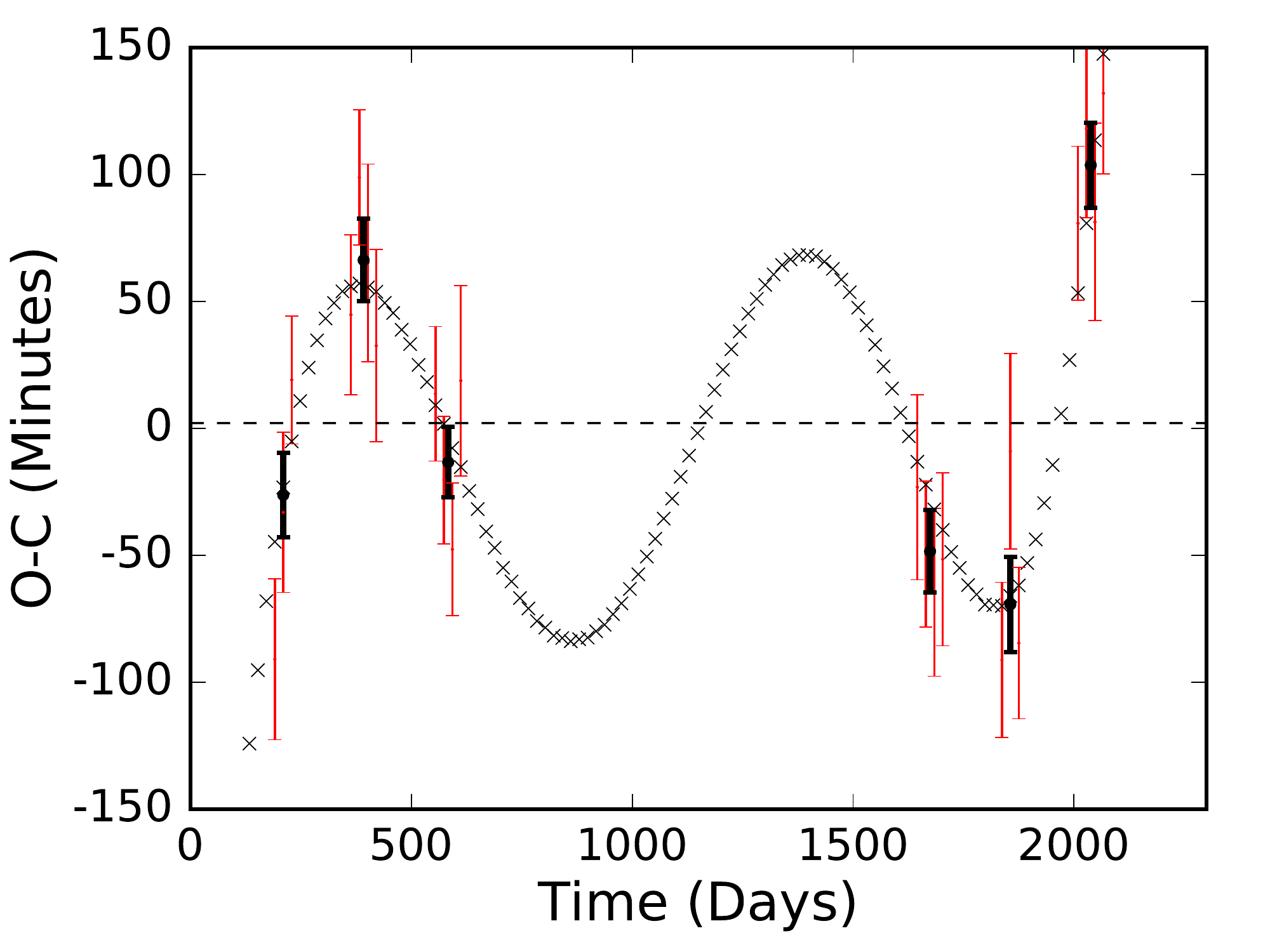}}
\caption{Simulated TTV signal from a two-planet system as observed by
  \WF\ (see Section \ref{sec:ttvs}).  Gray X labels correspond to the
  actual deviation form a linear ephemeris for each individual
  transit. For those observed during a simulated \WF\ season, typical
  uncertainties are added to the observed time of transit with data
  shown in red. The black points correspond to binned observations
  over an entire season. This hypothetical system would be confirmed
  by TTV observations in \WF\ data.}
\label{fig:ttv}
\end{figure}

First, we consider the precision with which \WF\, can measure the times of individual transits. We focus on the precision for the inner planet, because more transits will be observed over the course of the \WF\, mission. To begin, we fit a transit model to simulated transits for the inner planet, fixing the limb darkening to that predicted by \citet{Claret11} for a Sun-like star in the H-band but allowing the transit parameters to vary. After simulating many transits, we find a median uncertainty in the measured transit time for each individual transit of 28 minutes. Over the course of a single season, several transits will be observed. If we phase-fold over all observed transits inside an observing season, we find a median uncertainty in the average transit time of the folded transit of 15 minutes.

Given these expectations for the measurement precision, we then consider if \WF\, data can be used to identify interacting systems with TTVs. We use TTVFast \citep{Deck14} to integrate our test system as a dynamically interacting planetary system over a simulated \WF\, campaign. The simulated deviations of the transit times are shown as the gray X’s in Figure 6. We then simulate observations of these systems over hypothetical \WF\, seasons. For each observed transit, we add a random offset drawn from a uniform distribution on the range [25 minutes, 40 minutes], similar to the predicted scatter on measurements of the times of individual transits, and assign an uncertainty on the observed time of transit equal to this value (red points in Figure 6). The average time of transit for each particular season is also shown (black points). We purposefully schedule the \WF\, seasons to coincide with the smallest observed TTV signal to simulate a worst-case scenario. 

Fitting only the transit times of the inner planet, we find that a
dynamically interacting planet model fit the data considerably better
than a linear ephemeris ($\Delta \chi^2 = 64$).  In this case, these
planets would be easily confirmed via \WF\ observations. It is very difficult to contrive a set of observations of this planetary system that would not have detectable TTVs with \WF.  However, the inferred
masses of the transiting planets are a function of the unknown
eccentricity: a pair of 10 \mearth\ planets or a pair of 25
\mearth\ planets can both explain the observed TTVs.

Given this simulation, we conclude that \WF\ TTVs can be used to
confirm planets, but will not be robust for measuring their
masses. However, for the brightest stars, it may be possible to
identify particular transits that would 
be useful for precise determination of planet masses. 
These transits could then be targeted for observations from other facilities 
in order to measure their TTVs. Furthermore, based on demographics alone, this
method will be best for confirming smaller planets, which are more likely to
have companions to produce a TTV signal. In contrast to small planets, most
giant planets are most often detected in isolation, without an additional
transiting companion \citep{Steffen12c}. However, the detection of TTVs
requires the existence of a second planet.
So far, there is only one hot Jupiter
system with detected TTVs induced by the presence of an additional
planet \citep{Becker15}. We expect only a few of the hot
Jupiters detected by \WF\ will be confirmed by this method.

\subsection{Secondary Eclipses}

\begin{figure}[htbp!]
\centerline{\includegraphics[width=0.45\textwidth]{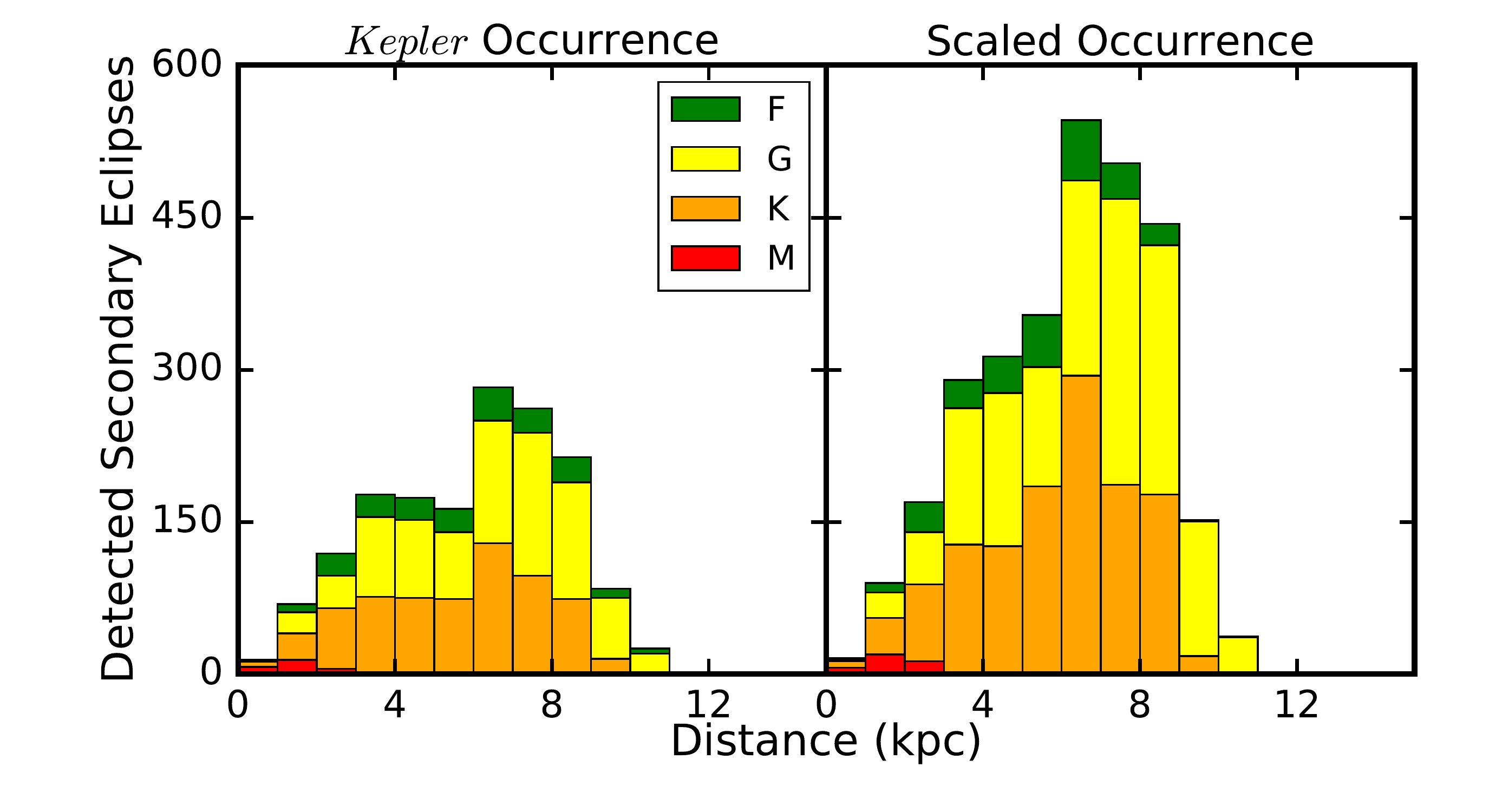}}
\caption{Simulated number of detected secondary eclipses of giant
  planets as a function of host star spectral type and distance.
  If planet occurrence is the same as the \kep\ field, we expect
  \WF\ to directly detect secondary eclipses of 1,700 planets.
  Assuming a local scaling with stellar metallicity, \WF\ will detect secondary eclipses of 2,900 planets. In both cases, the detected
  secondaries will predominantly be those of hot Jupiters transiting
  G and K dwarfs.}
\label{fig:sec_e}
\end{figure}

Although \kep\ shows that hot Jupiters are unlikely to exhibit TTVs, they can be
confirmed by observations of their secondary eclipses. The depth of
the secondary eclipse yields a measurement of the brightness
temperature, and thus the flux in that bandpass
\citep[e.g.][]{Charbonneau05}. Previous experience from \kep\ shows that with \kep\ data alone it is
difficult to confirm planets via secondary eclipses.  While
\kep\ detected thousands of planets, it was only able to confirm
planetary systems via detection of their phase curves and secondary
eclipses for a handful of these \citep[e.g.][]{Esteves13, Quintana13,
  Angerhausen15}.  Because the \kep\ bandpass spans approximately
$0.4-0.9$ $\mu$m, near the peak of a typical stellar spectrum but far
bluer than the typical planetary spectrum, only the hottest, largest
planets are detectable by their own emission.  However, \WF, with its
primary bandpass spanning 0.927-2.000 $\mu$m, will be significantly
more effective at observing planetary emission directly.

The bottom panels of Figure \ref{fig:HJtrans} show the secondary eclipses of a typical hot Jupiter, which has an eclipse depth equivalent to the transit of a $3 R_{\oplus}$ planet. As a more extreme example, WASP-12\,b's secondary eclipse depth integrated accros the \WF\ bandpass
is nearly 2 parts per thousand \citep{Croll11, Stevenson14b}, matching
the transit depth of a 4.9 \rearth\ planet.  Figure \ref{fig:Sensitivity} shows that we expect detections
of secondary eclipses of analogous planets to be detected around stars
as faint as $\textrm{W149} = 21.0$, by the end of the mission.  Even
smaller, cooler planets will be detected in secondary eclipse around
the million stars with W149 brighter than 15.0.

To determine the feasibility of observing secondary eclipses with \WF,
we model the secondary eclipses of each transiting planet injected in
Section \ref{ss:yield}.  We estimate the relative flux of each planet
and its host star in the \WF\ bandpass assuming the planet radiates as
a perfect blackbody.  This decision is a simplification of the
nonthermal physics of exoplanet atmospheres.  By comparing to
theoretical spectra from the BT-Settl spectral library of
\citet{Baraffe15} as well as actual near-IR observations of hot
Jupiters, we find it is too optimistic in the observed planet flux by
approximately a factor of two, so we divide all estimated fluxes by
that factor to account for this approximation.

We assume circular orbits for all planets, so that the duration of the
secondary eclipse is identical to the transit duration, and assume no
limb darkening or spatial variations in the received flux from the
planet itself.  We attempt to detect each secondary eclipse in the
same manner we detect each transit, declaring the eclipse detected if
it is observed at $7.1\sigma$, but making no $a priori$ assumptions
about the time of secondary in our search.

By the end of the mission, of the planets detected in our simulations
in Section \ref{ss:yield}, we expect to detect approximately 1,600
planets in secondary eclipse if the planet occurrence rate is the same as the
\kep\ field, and 2,900 planets if the planet occurrence rate scales
with metallicity in the same way as it does in the solar neighborhood.
As shown in Figure \ref{fig:sec_e}, these eclipses are predominantly
around G and K stars at a few kpc.  F stars are too luminous relative
to the planets for the secondary eclipses to be regularly detected,
while giant planets are too rare around M dwarfs to detect their
secondaries in significant numbers.  \WF\ will detect more secondary
eclipses than have been detected for all known exoplanets to date.
For many of these systems, \WF\ could plausibly detect 
phase curves as well, as we discuss in Section 5.3.1.

If secondary eclipses are to be used to confirm planetary systems,
observers will need to be able to separate true secondary eclipses from
false positive events.
It is unlikely that a background binary would be observed with a period as close
to that of a transiting system to mimic a secondary eclipse signal,
especially with a five-year time baseline.
The only conceivable false positive scenarios involve a single eclipsing binary
system masquerading as a transit and secondary eclipse.
As a planetary secondary eclipse depth will represent the eclipse of a $\sim 1000$ K body, most 
stellar-stellar secondary eclipses will be too deep to mimic a stellar-planetary
secondary eclipse. 
The most plausible false positive case is then a significantly blended 
stellar binary, blended either because of a bright background star, or because
the binary itself is in the background of the target star.

As we discuss in Section 5.1, because of the small pixel scale for \WF\ 
the rates of both of these events should be fairly low,
similar to \textit{Kepler} \citep{Morton11b}.
Even better, in many cases stellar-stellar secondary eclipses should be discernible from 
stellar-planetary eclipses as the durations of ingress and egress should be much longer
for stellar-sized objects than for planetary-sized objects. 
Additionally, high-resolution reconnaissance spectroscopy of these systems should
enable us to detect spectral signatures of blended stars in each system if they exist.
Therefore, the vast majority of candidate planetary secondary eclipse detections should be
real events, with a relatively small minority being astrophysical false positives.

\section{Validation of Transiting Planetary Systems}
\label{sec:validate}

If a planet cannot be directly confirmed, it is often still possible
to statistically validate the transit signal as caused by a planet at
a high degree of confidence.  \citet{Morton12} developed a method to
validate systems efficiently which has been used to validate more than
1,200 planets in the original \kep\ field. \citet{Morton15b} generalize this method to any field, which has enabled the
first catalogs of validated transiting planets with data from the
\textit{K2} mission \citep{Montet15b}.  If the potential transit
signal is actually caused by a brown dwarf or binary star system, there are several options for identifying
this.

\subsection{Blended Light}

Blended light from multiple stars in the PSF or in a given pixel crates two problems for reliably detecting transiting planets. First, the extra light can dilute a real planetary transit causing it to appear shallower than
otherwise expected.  Dilution could then lower the SNR of observed
transit signals, complicating the detection of small planets orbiting
faint stars. Second, the dilution can be so severe that a background eclipsing binary mimics a transiting
planet signal.

The \kep\ mission had smaller pixels on the detector relative to
wide-field, ground-based transit searches, making the stellar density
per pixel lower \citep{Morton11b}. As a consequence, it had a significantly lower rate of false positive
transit signals than previous transit search missions from the ground.

The Kepler Input Catalog contains approximately 4.5 million stars brigher than 
$K_p = 21$ that were observable by
the detector's 94.6 million pixel detector \citep{Haas10}, meaning on average
there were 4.8 stars for every 100 pixels. 
\WF\ has much smaller pixels than \kep\ (0.11" vs. 4"), but the campaign 
fields are significantly more crowded, containing approximately 300 million stars brighter than W149 = 28 \citep[see Figure 6 of ][ for an
  estimate of the stellar density as a function of
  magnitude]{Gould15}. As a result, the field will contain approximately 10 stars
  per 100 pixels within this limit, for a crowding rate approximately a factor of two larger than 
  \kep.
  The \kep\ limit is five magnitudes fainter than the limit for the faintest stars
  in the exoplanet survey ($K_p = 16$), where here the \WF\ limit is seven magnitudes
  fainter than the faintest planet hosts considered.
  Many of these potential blended stars will immediately be able to be ruled out as
  causes of false positive events, if their contribution to the light curve is less
  than the observed transit depth.
  The same techniques used to identify blends for
\kep\ and ground-based transit surveys should continue to be relevant
for \WF.  In particular, observations of centroid shifts of the
photocenter of light during transit events and differences in the
depth of alternating transits should provide information about false
positives.

\subsection{Z087 Photometry}

Transits of a dark object across the face of a star should be, to
first order, achromatic.  False positive events caused by eclipsing
binaries, where multiple objects are self-luminous, will have
wavelength-dependent depth variations as different portions of the
stellar SEDs are sampled at different bandpasses.  Multiband
photometry can then be used to separate transiting planets from
background eclipsing binary events.

In the \WF\ mission, one data point will be collected every 12 hours
in the Z087 filter, or one data point for every 47 obtained in W149.
For the example hot Jupiter transiting a
Sun-like star with a three-day period, only 24 data points will be
obtained during the transit event in Z087 over the entire mission,
approximately one data point for every six transits.  The situation
will be even worse for planets with longer orbital periods, or those
with higher impact parameters and shorter transit durations.

We can assume that the transit ephemeris and orbital parameters are
known from the W149 photometry used to detect planetary transit
signals. Therefore, we only need to fit three parameters in the Z087
transit model: two to describe the limb darkening and one to describe
the transit depth.  For this case, fitting the Z087 photometry we
measure a transit depth to a precision of 3.7\%. Therefore, an 11\%
difference in transit depth between Z087 and W149 is the minimum
detectable difference at $3\sigma$ confidence using data from the
entire mission. This is sufficient to rule out many, but not all,
stellar false positives.

For example, a false positive M7 dwarf with a temperature of 2900 K
and a radius equal to Jupiter's has a flux density smaller than the
Sun by a factor of 5.7 in the W149 filter and 11.6 in the Z087 filter,
leading to a 9\% change in the observed transit depth between the two
filters.  A moderate increase in the cadence of Z087 observations would be
required in order to detect these depth variations to identify false
positives.  However, as long as the orbit is aligned such that
secondary eclipses are observable from Earth, this star would induce a
2 ppt secondary eclipse, easily detectable with \WF\ photometry.

While Z087 photometry may be useful at the current cadence in
extreme cases, secondary eclipse photometry will be much more
significant, as long as the companion's orbit is aligned such that
secondary eclipses are visible.  An increased rate of $Z087$-band
photometry, perhaps as often as once every three hours, would provide
more opportunities to separate transiting hot Jupiters from
self-luminous brown dwarfs or giant planets.

Finally, we note that validation via wavelength dependent transit
depth can be complicated by the effects of starspots. This is true
both in the case where the planet crosses starspots, affecting the
light curve shape, and where starspots are located at different
latitudes, affecting the transit depth and out-of-transit flux.  Due
to the nature of the W149 bandpass, we expect spots to have a minimal
effect on the observed light curve.  They will be more prevalent in
the Z087 photometry, but still diminished relative to the
\kep\ bandpass.

{\subsection{Phase Curves}
\label{sec:PC}}

Although a transit is the most obvious signal in a light curve of a
planet orbiting a star, the companion planet affects the observed
light curve throughout its orbit.  Phase curve variations are the sum
of three separate effects: thermal emission, reflected light from the host star, 
relativistic Doppler
beaming, and ellipsoidal variations.  These variations have been
discussed in previous work as a method to measure planetary masses
\citep{Faigler11, Shporer11, Mislis12}, to detect new transiting objects
\citep{Faigler15b}, and to understand the atmospheres of transiting
planets \citep{Knutson07, Faigler15a}. 

\subsubsection{Thermal Emission}

A detection of a secondary eclipse is the detection of a planetary atmosphere.
At the moment before secondary eclipse ingress, the observed flux is the combined
flux from the star and the day side of the planet, while observations during the secondary
eclipse represent light from the star alone. 
If the planet has a large flux differential between its day side and night side, then we might
expect to see quasi-sinusoidal variations in the light curve over the course of each orbit
as the phase of the planet varies.
Phase curves have been observed for nearby transiting giant planets \citep{Knutson07}.
These enable a direct characterization of day-night temperature contrasts, rapid winds in
the atmospheres of these planets, and hot-spot offsets away from the substellar point of 
tidally locked systems \citep[e.g.][]{Knutson07, Knutson09, Stevenson14a}.

Here, we use WASP-43\,b \cite{Hellier11} as a test case to explore the detectability of
thermal emission of hot Juptiers with \WF.
\textit{HST} and \textit{Spitzer} phase curves of WASP-43
show the planet has a day-side temperature of 1700 K but a nightside temperature of 500 K
\citep{Stevenson16}. 
They also show that the peak in observed emission from WASP-43\,b occurs $40 \pm 3$ minutes 
before the secondary eclipse, or 
$12.3 \pm 1.0$ degrees east of the planet's substellar point. 
We analyze simulated \WF\ data to see if these signals would be detectable.

The day-night difference leads to an observable phase curve primarily in the mid-infrared, at
longer wavelengths than the \WF\ bandpass. 
From the emission spectrum of WASP-43b of \citet{Stevenson16}, the observed peak-to-peak amplitude of the
phase curve in the 4.5 $\mu$m \textit{Spitzer} bandpass is $3.99 \pm 0.14$ parts per thousand;
the expected signal 
integrated across the 
\WF\ bandpass is 0.4 parts per thousand, an order of magnitude smaller. 
This signal is larger than planetary signals that \WF\ will detect, but
while the secondary eclipse presents itself as a sharp ingress and egress separated by a few
hours, the phase curve signal is slowly varying over the course of the orbit. 
Its detection and characterization therefore requires any long-term systematics in the light curve to 
be well below the 400 part per million signal on few-day timescales.

If long-term systematics can indeed be maintained below the level of phase curve amplitudes in \WF,
then we should expect to be able to observe a phase curve for many of the systems for which we observe a 
secondary eclipse, especially if they are tidally locked.
We develop a simulated WASP-43\,b phase curve around a 15th magnitude star by creating a sinusoidal signal with
semiamplitude 200 parts per million and simulating observations, assuming the photometric uncertainty on each point is 1 part per
thousand. We then model the phase curve signal by fitting two Fourier modes to the data corresponding to observations of the dayside of the planet, following the method of \citep{Stevenson14a}. We fit the model and estimate
the uncertainties on the fit, leading to an uncertainty on the time of the brightest point, using the \textit{emcee} package 
\citep{Goodman10, Foreman-Mackey12}. From this fit, we measure an uncertainty of 3.7 degrees in our calculation of the
time of maximum, or 11 degrees at $3\sigma$. Formally, this would lead to a $3\sigma$ detection of a 12 degree offset
similar to the observed offset for WASP-43\,b, but detailed characterization of the planet's atmosphere would be limited.

Given our selection of planet, host star magnitude, and neglect of systematics, this represents a best-case scenario: the best cases could produce marginal detections of an offset at the 3$\sigma$ level, but more detailed characterization will not be possible. Therefore, it is likely that at best \WF\ phase
curves will represent a chance to probe day-night temperature contrasts of hot Jupiters with observed secondary eclipses,
but the prospects for a better understanding of these phase curves, such as detailed atmospheric transport, appears bleak.

\subsubsection{Reflected Light}

In addition to thermal emission, there is also an observable signal in reflected light from the host star, which often
dominates the phase curve signal in \textit{Kepler} data \citep{Shporer15}. For \WF\ we do not expect this signal to be
significant. The reflected light signal has amplitude
\begin{equation}
\frac{F_R}{F_0} = A_\lambda \bigg(\frac{2a}{R_p}\bigg)^2,
\end{equation}
where $F_R$ is the amplitude of the signal, $F_0$ the flux from the star, $A_\lambda$ the albedo, $a$ the orbital
semimajor axis and $R_p$ the planet radius.

While albedos are $\sim 0.1$ in the optical, averaged across W149 the
albedo for giant planets is typically $\sim 10^{-3}$. For the orbital parameters of WASP-43\,b, we would then expect the
amplitude of the signal to be $\sim 10^{-7}$, well below what is observable. Typical albedos in the Z087 bandpass will
be on the order of $10^{-2}$, so the amplitude of the signal will be an order of magnitude larger, but as the number of
observations in this bandpass will be a factor of 50 lower, negating much of this benefit and making systematics harder
to identify and mitigate.

\subsubsection{Doppler Beaming}

As a planet and host star orbit their mutual center of mass, the flux emitted from the star is
beamed towards the direction of travel due to the changing
the velocity of the host star.  A consequence of special
relativity, the signal is observable at the non-relativistic speeds at
which stars move during their orbits.  To first order, the amplitude
of the beaming signal is
\begin{equation}
\frac{F_D}{F_0} = (3-\alpha)\frac{K_s}{c}
\end{equation}
where $F_D$ is the amplitude of the signal, $F_0$ the flux from the
stationary star, $\alpha$ the shape of the SED at the observed
wavelength, $K_s$ the Doppler semiamplitude of the star, and $c$ the
speed of light \citep{Loeb03}.  The SED is relevant because, as the
star's velocity is modulated, the Doppler shift affects what features
of the stellar spectrum fall in our bandpass.  For most stars, the
W149 filter will fall on the Rayleigh-Jeans tail of the SED, where
$\alpha = 2$.

For a typical hot Jupiter ($K_s \sim 150$ m s$^{-1}$), the beaming amplitude
will be $\sim 0.5$ parts per million, well below the sensitivity of
\WF.  However, this effect will be useful for detecting more massive
objects of similar radii masquerading as hot Jupiters, such as brown
dwarfs or very low mass stars.  A $50$ \mjup\ object with a three-day
period would exhibit a 25-ppm signal. In Section \ref{sec:transits}
we determined that we can measure a transit depth to a precision of $40$
ppm. That transit event has a duration of 1.5 hours, whereas the
beaming signal occurs throughout the orbit.
This implies that beaming will be measured
in many of these false positive scenarios.
However, the smooth, sinusoidal signature of the beaming signals may be harder
to distinguish from instrumental noise than the sharp box-shaped signature
of a transit.
Moreover, if the secondary is luminous in the \WF\ bandpass, it will exhibit its
own Doppler beaming signal which will destructively interfere with the primary
signal, reducing the magnitude of the observable effect \citep{Shporer10}.

\subsubsection{Ellipsoidal Variations}

Ellipsoidal variations are an achromatic phenomenon caused by changes
in the sky-projected shape of a star as a planet orbits, affecting the
star's gravitational potential.  The signal has twice the frequency of
the planet's orbit.  Following \citet{Loeb03}, to first order the
magnitude of the signal is
\begin{equation}
\frac{F_E}{F_0} \sim \beta \frac{M_p}{M_s} \bigg(\frac{a}{R_s}\bigg)^{-3},
\end{equation}
Here, $\beta$ is a term which depends on the nature of gravity
darkening for the host star.  For Sun-like stars, this value is
approximately 0.45. $M_p/M_s$ is the mass ratio between the planet and
star and $a/R_s$ is the reduced orbital semimajor axis.

In general, the signal is of a similar magnitude to the Doppler
beaming signal, and only likely to be useful in separating brown
dwarfs from planets: transiting planets will only be notable by a
nondetection of their ellipsoidal variations.

\citet{McDonald14} note that in the case of \textit{Euclid}, a
color-dependence in observed ellipsoidal variations would be a
signature of a background eclipsing binary, as the signal would be
achromatic but the relative flux between the foreground and background
target would vary between the two bandpasses.  The same is true here,
although with the cadence of Z087 observations we do not expect this
effect to be detectable.  In any cases where such an effect would be
detectable, variations in the eclipse depth between the bandpasses
would also be detectable, likely at a much higher significance.
Unlike the Doppler beaming case, because the signal occurs at twice
the orbital period, additional signals from a luminous secondary
would constructively interfere with the signal from the primary,
making the signal even easier to detect.

\subsection{Ground-based followup}
Ground-based adaptive optics imaging is typically used to rule out
false positive blends of nearby star systems and to understand the
level of dilution in transit light curves \citep{Law14}.  In
principle, the transiting planet candidates discovered by \WF\ can be
followed up by adaptive optics (AO) systems on 10-meter telescopes,
or upcoming 30-meter class
telescopes that are expected to be built before the \WF\ launch date.  
These observations may not provide much leverage over the
\WF\ data themselves.  The diffraction limit of a 30-meter telescope
in $K$-band is $\sim 20$ milliarcseconds.  While considerably smaller
than the \WF\ pixel scale of 0\farcs11/pixel, this still corresponds
to a projected separation of $\approx$20 au for a Sun-like star with
W149 = 14.5, at a distance of $\approx$1 kpc.  The diffraction limit
also corresponds to a projected separation of 200 au for a Sun-like
star at 1 kpc with W149 = 19.5, meaning many bound binary companions
will be unresolved even when operating a thirty-meter telescope at the
diffraction limit.

\section{Galactic Exoplanet Demographics}

\begin{figure}[htbp!]
\centerline{\includegraphics[width=0.5\textwidth]{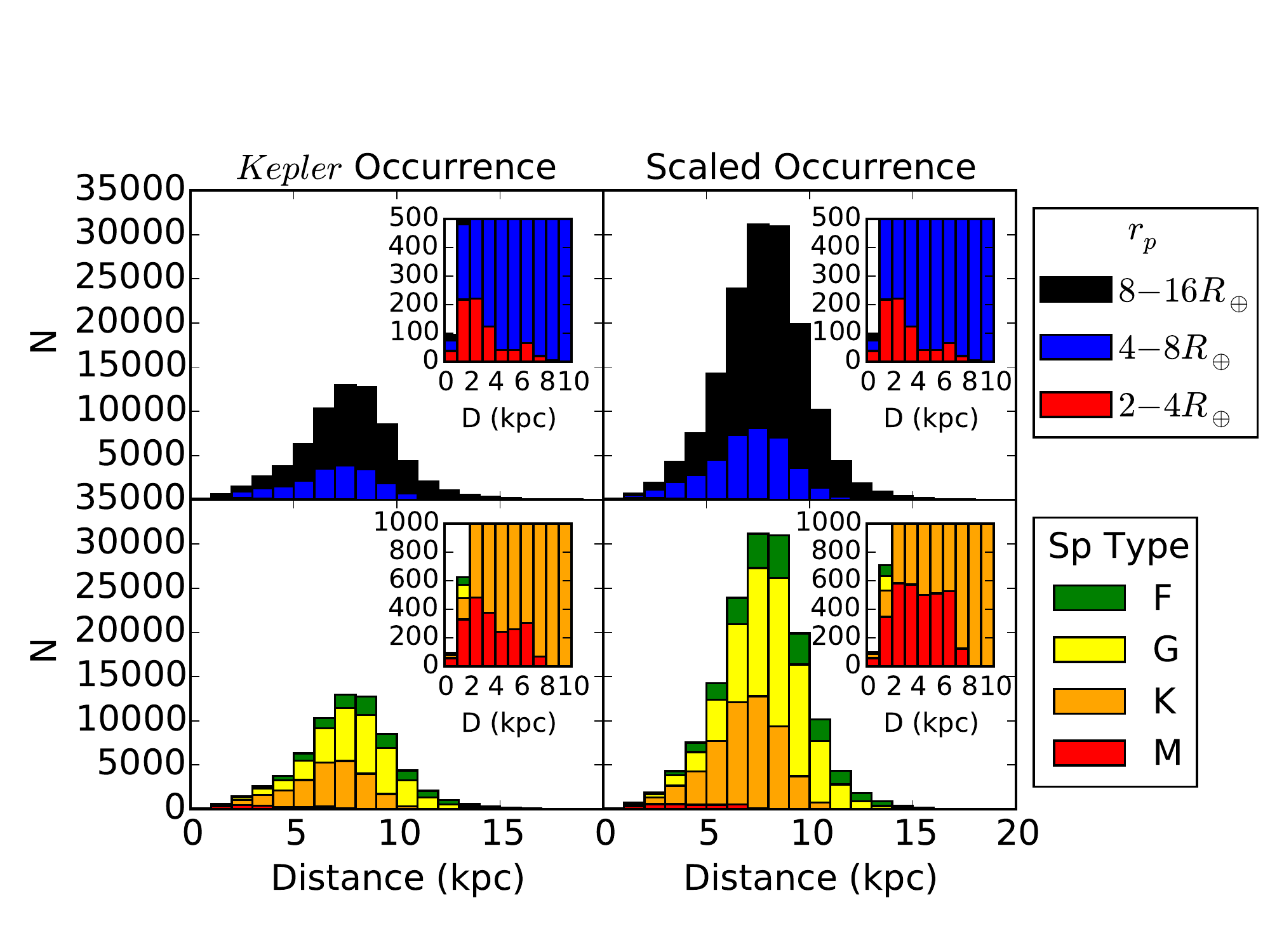}}
    \label{fig:pl_v_dist}
    \caption{The distribution of simulated transiting planets as a
      function of galactic distance. The left panels assume the planet
      occurrence rate is identical to that of \kep; the right panels
      assume the occurrence rate scales with metallicity. The top
      panels show the breakdown of planet discoveries as a function of
      planet radius, and the bottom panels show the breakdown as a
      function of host spectral type. The insets highlight the
      distributions for small planets (top) and small stars (bottom)
      for which the total numbers are comparatively small.}
\end{figure}

With the simulation described in Section \ref{ss:yield}, we are also
able to evaluate the number of transiting planets that will be
discovered by \WF\ as a function of distance from the Sun. 
\citet{Gould15} contains a detailed discussion of
measuring parallaxes with \WF, including a discussion of the potential
sources of systematic errors. In summary, \WF\, will
have a single measurement, astrometric precision of 0.7 mas for stars
as faint as $H_{AB}=19.6$ and 1.7 mas for $H_{AB}=21.6$
\citep{Spergel15}. Given that there will be 40,000 measurements of
each star over the course of the mission, in the absence of severe
systematics, \citet{Gould15} determine \WF\, should measure the relative
parallaxes of all planets considered in this study. 
These relative
parallaxes can also be tied to the {\em Gaia} system to determine the
absolute parallaxes. 

In real data, determining the distances to stars may be
more difficult.
We will not understand each pixel perfectly, so there should
be underlying systematic effects. 
A reasonable comparison case may be the astrometry achieved by using
the spatial scan mode of \textit{HST} \citep{Riess14}. 
This method leads to an astrometric precision of 20-40 $\mu$as.
The pixels of
\WF\ are a factor of two larger, which would lead to a factor of two larger
uncertainty on the parallax. 
If the systematics can be controlled to this level (i.e. 80 $\mu$as), then \WF\ can measure 
$3\sigma$ parallaxes out to 4 kpc and place upper limits beyond that. It should then be possible 
to construct a sample of known disk stars with measured parallaxes and a sample of probable bulge 
stars with parallaxes consistent with zero.

Even with no distance measurements, it will still be possible to investigate 
the Galactic distribution of hot and warm planets. The different Galactic stellar populations 
have different proper motion distributions. WFIRST's astrometric precision will be easily 
good enough to measure accurate proper motions for each planet host, and with a large 
population of planet hosts it will be possible to select large, statistically clean samples 
of bulge and disk planets in the same way that is currently done for stars \citep{Clarkson08}.

Figure \ref{fig:pl_v_dist} shows the distribution of planets from our
simulation as a function of distance from the Sun. These planets are
distributed all along the line of sight, including into the bulge of the
galaxy. Of particular interest are planets that can be directly confirmed via
secondary eclipses. Their distribution is shown in Figure \ref{fig:sec_e}.
Given that \WF\ should detect thousands of confirmed hot Jupiters, it will be
possible to study their distribution towards the center of the galaxy. This
measurement will test the variation in occurrence rate of short-period giant
planets in the disk and the bulge, and whether or not planet formation is
suppressed in the bulge as suggested by \citet{Thompson13}.

\section{Conclusions}

While ostensibly a microlensing mission, \WF\ will provide a
tremendous opportunity for the study of short-period, transiting
planets as well. We have shown in Section \ref{ss:yield} that if the
occurrence rate of planets is the same as for the main \kep\ field,
\WF\ could detect 70,000 transiting planets with sizes as small as $2
R_{\oplus}$ at distances of up to 10 kpc or more. If the occurrence rate scales with metallicity as in
\citet{Johnson10a}, we expect as many as 150,000 planets, as the
\WF\ field is more metal-rich than the solar neighborhood. All of these systems should have measured parallaxes \citep{Gould15}.

While the vast majority of these planets will be found around stars
too faint for followup observations, we explore various options for
confirming or validating these planets directly from the \WF\ data. We
find that secondary eclipse depth measurements can be used to confirm
as many as 2,900 giant planets, which can be detected at distances of
$>8\,$ kpc. From these confirmed \WF\ planets, we will be able to
measure the variation in the occurrence rate of short-period giant
planets. Furthermore, we show that \WF\ is capable of detecting
transit timing variations which can be used to confirm the planetary
nature of some systems, especially those with smaller planets.

The transiting planets found by \WF, especially those that can be confirmed,
will provide unprecedented information about how planetary system architectures
vary with galactic environment. Although the transiting planets and \WF\
microlensing planets will generally not be found around the same host stars,
both samples probe the same planetary population but at very different
planetary separations. Figures 14 and 15 of \citet{McDonald14} summarize the complementarity of the two techniques in the context of {\em Euclid}. It is clear that by combining the two samples of planets, we can probe planetary system architecture from very small separations to beyond 10 au.

\acknowledgements

We thank Avi Shporer (JPL) for comments which improved the quality of this manuscript.

Work by B.T.M., J.C.Y., and M.T.P. was performed in whole or in part under contract with
the California Institute of Technology (Caltech)/Jet Propulsion
Laboratory (JPL) funded by NASA through the Sagan Fellowship Program
executed by the NASA Exoplanet Science Institute. This research has
made use of the NASA Exoplanet Archive, which is operated by the
California Institute of Technology, under contract with the National
Aeronautics and Space Administration under the Exoplanet Exploration
Program.

\end{document}